\documentclass[usenatbib]{mn2e}
\usepackage{natbibmnfix,graphicx,times}

\newcommand{\Mpc}{\mbox{ Mpc}}

\newcommand{\secinv}{\mbox{ s$^{-1}$}}

\newcommand{\Hz}{\mbox{ Hz}}

\newcommand{\Msun}{\mbox{ M$_\odot$}}

\newcommand{\hunits}{\mbox{ km s$^{-1}$ Mpc$^{-1}$}}

\newcommand{\bxhi}{\bar{x}_{\rm HI}}
\newcommand{\xhi}{x_{\rm HI}}

\newcommand{\lya}{Ly$\alpha$ }
\newcommand{\lyans}{Ly$\alpha$} % Use this version if Ly\alpha is
\newcommand{\bq}{\begin{equation}}
\newcommand{\eq}{\end{equation}}
\newcommand{\bqa}{\begin{eqnarray}}
\newcommand{\eqa}{\end{eqnarray}}
\def\VEV#1{\left\langle #1\right\rangle} % This is \VEV{x} => <x>

\newcommand\lsim{\mathrel{\rlap{\lower4pt\hbox{\hskip1pt$\sim$}}
        \raise1pt\hbox{$<$}}}
\newcommand\gsim{\mathrel{\rlap{\lower4pt\hbox{\hskip1pt$\sim$}}
        \raise1pt\hbox{$>$}}}
\newcommand{\taudamp}{\tau_D}
\newcommand{\colden}{\mbox{ cm$^{-2}$}}
\newcommand{\lobs}{\lambda_{\rm obs}}

\newcommand{\qnamefourtwo}{J1148+5251}
\newcommand{\qnametwoeight}{J1030+0524}
\newcommand{\qnametwotwo}{J1623+3112}

\title[The \lya damping wing]{\lya damping wing constraints on inhomogeneous reionization}

\author[Mesinger \& Furlanetto]{Andrei Mesinger \& Steven R.  Furlanetto \\ Yale Center for Astronomy and Astrophysics, Yale University, New Haven, CT 06520
}

\begin{document}

\maketitle

\begin{abstract}
One well-known way to constrain the hydrogen neutral fraction, $\bxhi$, of the high-redshift intergalactic medium (IGM) is through the shape of the red damping wing of the \lya absorption line.  We examine this method's effectiveness in light of recent models showing that the IGM neutral fraction is highly inhomogeneous on large scales during reionization.  Using both analytic models and ``semi-numeric" simulations, we show that the ``picket-fence" absorption typical in reionization models introduces both scatter and a systematic bias to the measurement of $\bxhi$.  In particular, we show that simple fits to the damping wing tend to \emph{overestimate} the true neutral fraction in a partially ionized universe, with a fractional error of $\sim 30\%$ near the middle of reionization.  This bias is generic to any inhomogeneous model.  However, the bias is reduced and can even underestimate $\bxhi$ if the observational sample only probes a subset of the entire halo population, such as quasars with large HII regions.  We also find that the damping wing absorption profile is generally steeper than one would naively expect in a homogeneously ionized universe.  The profile steepens and the sightline-to-sightline scatter increases as reionization progresses.  Of course, the bias and scatter also depend on $\bxhi$ and so can, at least in principle, be used to constrain it.  Damping wing constraints \emph{must} therefore be interpreted by comparison to theoretical models of inhomogeneous reionization.
\end{abstract}


\section{Introduction} \label{intro}

The reionization of hydrogen in the intergalactic medium (IGM) is a landmark event in the early history of structure formation, because it defines the moment at which galaxies (and black holes) affected every baryon in the Universe.  As such, it has received a great deal of attention -- both observationally and theoretically -- in the past several years.  Unfortunately, the existing observational evidence is enigmatic (see \citealt{fan06-review} for a recent review).  Electron scattering of cosmic microwave background photons implies that reionization occurred at $z \sim 10$, albeit with a large uncertainty \citep{page06}.  On the other hand, \lya forest spectra of  quasars at $z \sim 6$ show some evidence for a rapid transition in the globally-averaged neutral fraction, $\bxhi$ (e.g., \citealt{fan06}).  However the \lya absorption is so saturated in the \citet{gunn65} trough (with optical depth $\tau_{\rm GP} \ga 10^5 \bxhi$) that constraints derived from that spectral region \citep{fan06, maselli07} are difficult to interpret (e.g, \citealt{lidz06, becker07}).  

Another probe is the red damping wing of the IGM \lya absorption:  the line is so saturated at these redshifts that even photons that are emitted redward of the \lya resonance can suffer significant absorption from the strong damping wings of that transition.  This has a number of consequences for high-redshift observations.  

For example, surveys that search for high-$z$ galaxies through their \lya emission lines will find fewer and fewer galaxies as the IGM becomes more and more neutral \citep{haiman02-lya, santos04}, although galaxy clustering strongly moderates this decline \citep{furl04-lya, furl06-lya, mcquinn07, mesinger07-lya}.  Such surveys have now detected objects at $z \sim 6.5$--$9$ (e.g., \citealt{kashikawa06, iye06, stark07}), but their implications for reionization are unclear \citep{malhotra04, haiman05-lya, malhotra06, kashikawa06, dawson07, dijkstra07, mcquinn07-lya, mesinger07-lya}.  

The evolution of galaxy abundances and clustering measures the damping wing absorption in a statistical sense, but even more information can potentially be gleaned from the damping wing absorption profiles in individual objects \citep{miralda98}.  For the galaxies described above, this information is difficult to extract because of their faintness and the complicated origins of their \lya emission lines \citep{mcquinn07-lya}.  

However, high signal-to-noise spectra of bright objects could be extremely helpful. If the damping wing profile from IGM absorption can be isolated from these spectra, this would provide detailed information on the neutral gas along each particular line of sight (LOS) -- rather than the statistical information available from most other probes.  This is very useful, as reionization is expected to be highly inhomogeneous.

There are two candidates for such high signal-to-noise spectra at high-redshifts:  quasars and gamma-ray bursts (GRBs).  Quasars present several challenges: complicated intrinsic spectra, biased IGM environments \citep{barkana04-grb, lidz07}, and large HII regions (which significantly weaken the damping wing absorption redward of the quasar \lya line, and can necessitate detailed spectral analysis of the blue side of the line; \citealt{madau00, mesinger04-mockprox}).  Nevertheless, there have already been two claims of damping wing detections in high-redshift spectra, both using quasars from the Sloan Digital Sky Survey (SDSS).  \citet{mesinger04} detected a $\bxhi \gsim 0.2$ damping wing through the decreased fluctuations in the total Ly$\alpha$ optical depth near the edge of the HII region surrounding \qnametwoeight\ ($z_S=6.28$).  Similarly, by simulating the optical depth distributions blueward of the Ly$\alpha$ line center and comparing them with deep observations, \citet{mesinger07-prox} detected the presence of a $\bxhi \gsim 0.033$ damping wing in the spectra of \qnametwoeight\ and \qnametwotwo\  ($z_S=6.22$).  The maximum likelihood was at $\bxhi=1$ for both quasars.

The second set of candidates, GRBs, have fewer obstacles to overcome.   Long-duration GRBs are believed to be remnants of massive stars (and so trace the bulk of the star formation, which probably occurs in lower-mass halos with more ``typical" IGM environments), and their afterglows have extremely simple power-law intrinsic spectra (see, e.g., \citealt{piran05} for a review).  The event rates at high redshifts may be quite high, and cosmological time-dilation helps to identify the sources when they are still bright \citep{bromm02-grb, ciardi00, lamb00, mesinger05}.  As a result, there is a great deal of optimism in the literature regarding their potential for damping-wing measurements (e.g. \citealt{miralda98, barkana04-grb}).  The highest-redshift GRB afterglow observed so far (at $z \approx 6.3$), has already been used to constrain the global neutral fraction at that time \citep{kawai06, totani06}.  Unfortunately, this object illustrates the major difficulty with the red damping wing test for GRBs:  intrinsic absorption in the host galaxies \citep{miralda98}.  Most GRBs are now known to have large columns of associated neutral hydrogen \citep{vreeswijk04, chen04-grb}. Roughly $20\%$ of well-studied objects have $N_{\rm HI} \la 10^{20} \colden$ \citep{chen07}, although nearly all of the objects in this sample are at $z \la 6$.   

The $z \approx 6.3$ GRB does appear to have intrinsic absorption with $N_{\rm HI} \sim 10^{21.6} \colden$ \citep{totani06}, which makes it difficult to constrain the IGM absorption.  In principle, it is still possible because isolated HI absorbers have different spectral profiles than the IGM (with the optical depth inversely proportional to the wavelength offset squared for isolated absorbers, and to the wavelength offset itself for the IGM).  The two sources can then be separated by looking at the shape of the absorption.  \citet{totani06} found a best fit with $\bxhi=0$ and estimated that $\bxhi \la 0.17$ ($0.60$) at 68\% (95\%) confidence.  Better constraints will require faster followup (when the afterglow is brighter) and systems with less intrinsic absorption.

To date, the red damping wing test has generally been assumed to be simple and straightforward.  It is usually argued that the absorption is sensitive to a large path length in the IGM, so that small-scale clumpiness can be ignored and that the ionized fraction can be taken to be uniform (for an exception, see \citealt{barkana02}).  However, most models of reionization have much more inhomogeneous distributions of neutral and ionized gas, with discrete HII regions surrounding clusters of galaxies, and a sea of nearly neutral gas separating them (e.g., \citealt{arons72, shapiro87}).  Such a picture is inevitable when hot stars ionize the gas.  Moreover, the most recent models show that the ionized bubbles can become quite large even relatively early in reionization, with sizes $\ga 10 \Mpc$ when $\bxhi \sim 0.5$ \citep{furl04-bub, furl05-charsize, iliev05-sim, zahn07-comp, mcquinn07, mesinger07}.  

Because the damping wing is sensitive to fluctuations on Mpc scales, it is actually not a good approximation to take the IGM ionized fraction to be constant.  In this paper, we will examine whether (and how) the damping wing can actually be used to constrain the reionization process.  We summarize the basic physics of the line in \S \ref{lya}.  We then examine a series of toy models of the ``picket-fence" absorption typical of the IGM during reionization in \S \ref{toy}.  In particular, we show that interpreting measurements with the naive view of a uniform IGM is not only subject to significant scatter (from the different networks of ionized bubbles intersected along different lines of sight) but also a substantial systematic bias.  In \S \ref{sim}, we describe the ``semi-numeric" simulations used to generate our main results, which we present in \S \ref{results}.  This more detailed picture confirms that scatter between different lines of sight and bias relative to the naive view will be critical in interpreting any observed sources.  Finally, we conclude in \S \ref{disc}.

When this project was nearing completion, we learned of a similar effort by \citet{mcquinn07-damp} and refer the reader there for a complementary discussion.

In our numerical calculations, we assume a cosmology with $\Omega_m=0.26$, $\Omega_\Lambda=0.74$, $\Omega_b=0.044$, $H=100 h \hunits$ (with $h=0.74$), $n=0.95$, and $\sigma_8=0.8$, consistent with the most recent measurements \citep{spergel06}.  Unless otherwise specified, we use comoving units for all distances.

\section{The \lya Damping Wing} \label{lya}

We compute the total \lya optical depth at an observed wavelength $\lobs=\lambda_\alpha(1+z)$ along a line of sight (LOS) centered on a halo at $z_S$.  We do this by summing the damping wing optical depth, $\taudamp$, contribution from each neutral hydrogen patch (extending from $z_{bi}$ to $z_{ei}$ for the $i$th patch, with $z_{bi}>z_{ei}$) encountered along the LOS, using the approximation \citep{miralda98}:
\bqa
\taudamp(z) & = & {\tau_{GP} R_\alpha \over \pi} \sum_{i} \left\{ \xhi(i) \left( {1+z_{bi} \over 1 + z} \right)^{3/2} \right. \nonumber \\
& & \times \left. \left[ I\left( \frac{1+z_{bi}}{1+z}\right) - I\left( \frac{1+z_{ei}}{1+z}\right)  \right] \right\}
\label{eq:jordi_tau}
\eqa
where $\tau_{GP} \approx 7.16 \times 10^5 [(1+z_S)/10]^{3/2}$ is the \citet{gunn65} optical depth of the IGM in our cosmology, $R_\alpha=\Lambda/(4 \pi \nu_\alpha)$, $\Lambda=6.25 \times 10^8 \secinv$ is the decay constant for the \lya resonance, and $\nu_\alpha=2.47 \times 10^{15} \Hz$ is the rest frequency of the \lya line.  Finally,
\bq
I(x) \equiv \frac{x^{9/2}}{1-x} + \frac{9}{7} x^{7/2} + \frac{9}{5} x^{5/2} + 3x^{3/2} + 9x^{1/2} - \ln \left| \frac{1+x^{1/2}}{1-x^{1/2}} \right| ~ .
\label{eq:idefn}
\eq
This expression is only valid far from line center, but that is acceptable because the optical depth is so large (and therefore unmeasurable) at line center anyway.  It also assumes $\Omega_m(z)=1$, which is an excellent approximation at the high redshifts relevant here.

For the remainder of this paper, we will assume that $\xhi(i)=1$:  in other words, the neutral patches between ionized zones are indeed fully neutral.  This is an excellent approximation in numerical simulations of reionization by hot stars (although less good if X-rays contribute substantially).  

For the analytic calculations in the following section, it is useful to approximate $I(x)$ by its asymptotic limit when $|x-1| \ll 1$.  In that limit, equation~(\ref{eq:jordi_tau}) can be written
\bq
\taudamp(z) \approx {\tau_{GP} R_\alpha \over \pi} {(1+z) c \over H(z)} \sum_{i} ( R_{bi}^{-1} - R_{ei}^{-1} ) ,
\label{eq:tau-approx}
\eq
where $R_{bi}$ and $R_{ei}$ are the comoving distances from redshift $z$ to redshifts $z_{bi}$ and $z_{ei}$, the beginning and end of the $i$th neutral patch in redshift space.  Note that we have further assumed $z_{ei} \approx z_{bi} \approx z$ (which is an excellent approximation in most of the cases of interest).  Although this asymptotic form is not accurate enough for detailed calculations or inferences from observations, it contains all of the essential features of the damping wing solution and so is useful to understand the source of many of the effects we will describe.  We will also use wavelength units, $\Delta \lobs = \lobs - \lambda_\alpha (1+z_{bi})$.  

The $R_{bi}^{-1} \sim (\lobs/\Delta \lobs)$ decline of equation~(\ref{eq:tau-approx}) is much gentler than a damped \lya absorber (DLA) at the same location (which falls off like $[\lobs/\Delta \lobs]^2$).  This is because of the large sizes of the IGM damping wing absorbers:  at large wavelength offsets, a longer path length is able to contribute, which moderates the decline \citep{miralda98}.  It is this property that (one hopes) will allow us to distinguish IGM absorption from neutral gas within the host galaxies of GRBs, for example.  

A common alternative approach to ours (where we have explicitly broken up the LOS into ionized and neutral patches) is to assume that the damping wing averages over a sufficiently long path length so that the sum over the patches can be replaced by an average neutral fraction, $\bar{x}_D$:
\bq
\taudamp(z) \approx {\tau_{GP} R_\alpha \over \pi} \bar{x}_D \left( {1+z_{b1} \over 1 + z} \right)^{3/2} \left[ I\left( \frac{1+z_{b1}}{1+z}\right) - I\left( \frac{1+z_{e}}{1+z}\right)  \right],
\label{eq:xhd}
\eq
or in the asymptotic limit of $I(x)$, in analogy with equation~(\ref{eq:tau-approx}),
\bq
\taudamp(z) \approx {\tau_{GP} R_\alpha \over \pi} {(1+z) c \over H(z)} \bar{x}_D ( R_{b1}^{-1} - R_{e}^{-1} ) ,
\label{eq:xhd-tau-approx}
\eq
where $z_{b1}$ and $R_{b1}$ denote the edge of the closest neutral gas and $z_e$ and $R_e$ denote the largest distance at which neutral gas sits (the result is, however, quite insensitive to $z_e$, as long as $z_e \not\approx z_{b1}$).  In this simple picture, the spectral profile of the absorption is well-known.  

The basic measurement is then to extract $z_{b1}$ (or $R_{b1}$) and $\bar{x}_D$ from a fit to $\taudamp(z)$.  It is commonly assumed in the literature that $\bar{x}_D$ will be an accurate and unbiased estimator of $\bxhi$; in the remainder of this paper, we will critically examine these expectations.  Of course, in principle a much more sophisticated fit may be performed with many $R_{bi}$ and $R_{ei}$.  However, in practice the dependence on any individual element (other than $R_{b1}$) is small, and adding more parameters will rapidly weaken constraints from the fit.

\section{Some Illustrative Toy Models} \label{toy}

We begin by constructing a series of toy models, using the approximate form of the damping wing optical depth in equation~(\ref{eq:tau-approx}), that will show the basic features of the measurement.  For this simple case, we will define the \emph{apparent} neutral fraction $\bar{x}_D$ by equating the right hand sides of equations (\ref{eq:tau-approx}) and (\ref{eq:xhd-tau-approx}) and taking the $R_e \rightarrow \infty$ limit in the latter:
\bq
\bar{x}_D \equiv R_{b1} \VEV{\sum_{i} ( R_{bi}^{-1} - R_{ei}^{-1} )}.
\label{eq:xhd-approx}
\eq
where we have assumed that the only second parameter measurable from the absorption profile is the location of the near edge of neutral gas, $R_{b1}$.\footnote{Note that independently measuring $R_{b1}$ from the spectrum becomes non-trivial for large $R_{b1}$, as the resonant absorption from residual HI within $R_{b1}$ can become strong enough to wipe out detectable flux \citep{mesinger04-mockprox, bolton07}.}  We will also assume that the summation extends to infinity, although this does not affect our conclusions.

\subsection{Bias in the Inferred Neutral Fraction} \label{bias}

To begin, we consider a model where the IGM is divided into ionized and neutral patches of fixed lengths $R_b$ and $f R_b$, respectively, where $f \equiv \bxhi/(1-\bxhi)$ ensures that the mean neutral fraction along the given line of sight is $\bxhi$.  For all of our toy models, we will assume that the source sits in the middle of its host bubble, so $R_{b1}=R_b/2$; keep in mind, however, that this does not affect our results because we have assumed that $R_{b1}$ is independently measured from the data.  In reality, more sophisticated techniques may be needed to constrain $R_{b1}$ (e.g. \citealt{mesinger04, mesinger07-prox}).  Equation~(\ref{eq:xhd-approx}) then becomes
\bqa
\bar{x}_D & = & {1 \over 2} \sum_{k=1}^{\infty} \left[ {1 \over (k-1/2) + (k-1)f} - {1 \over (k-1/2) + kf} \right] \\
& = & \pi (1 - \bxhi) \cot \left[ {\pi (1 - \bxhi) \over 2} \right].
\label{eq:bias-fix}
\eqa

\begin{figure}
\begin{center}
\resizebox{8cm}{!}{\includegraphics{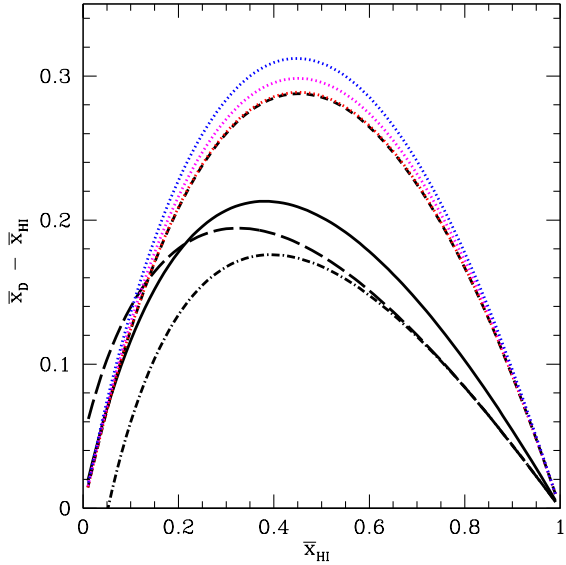}}\\%
\end{center}
%\plotone{dampingBias.eps}
\caption{The damping wing bias that results from assuming a constant neutral fraction throughout the IGM.  Each curve shows the difference between the inferred and true neutral fractions.  The short-dashed curve assumes that ionized and neutral patches have fixed sizes.  The thin dotted curves assume the same but take the full damping wing profile, for $R_b=1,\,10,$ and $30 \Mpc$, from bottom to top.  The solid and long-dashed curves fix $R_b$ but assume respectively that the neutral patches are uniformly distributed in size between $[0,\,2 f R_b]$ and exponentially distributed (with the same expectation value).  The dot-dashed curve assumes that both ionized and neutral patches are exponentially distributed.}
\label{fig:bias-toy}
\end{figure}

The short-dashed line in Figure~\ref{fig:bias-toy} shows the difference $(\bar{x}_D - \bxhi)$ for this model as a function of the true average neutral fraction, $\bxhi$.  It is obviously \emph{not} a particularly good estimator.  The error peaks at $\sim 0.3$ when $\bxhi \sim 0.5$ (although note that the fractional error continues to increase as $\bxhi \rightarrow 0$).  We see that $\bar{x}_D$ \emph{overestimates} the neutral fraction, because the nearest material dominates the absorption.  In the ``true" model, this material is always fully neutral and so absorbs considerably more radiation than would be expected in a simple, uniformly ionized model.

%%%%%%%%%%%%%%%%% FIGURE 1

The thin dotted curves in Figure~\ref{fig:bias-toy} show explicitly that this bias does not depend on our use of the approximate equation~(\ref{eq:tau-approx}).  Here we take the full expression for $\tau_D(z)$ (i.e. we estimate $\bar{x}_D$ by equating the right hand sides of equations \ref{eq:jordi_tau} and \ref{eq:xhd}) but consider the same sequence of ionized and neutral regions.  The three curves assume $R_b=1,\,10,$ and $30 \Mpc$, from bottom to top.  All also only include neutral patches at $z>6$ (with the source at $z_s=9$).  Obviously, the $1/R$ model is an excellent estimate.  Interestingly, the bias is only a weak function of the actual size of the patches, although it does increase slowly with $R_b$.  This suggests that the bias can probably be studied fairly robustly.  

The solid line represents a slightly more sophisticated model.  Here we keep the ionized patches at a fixed size $R_b$ but draw the length of each neutral patch from a uniform distribution over the range $[0,  \, 2 f R_b]$; thus the mean neutral fraction is still $\bxhi$, but different lines of sight construct it in different ways.  In this case, 
\bq
\bar{x}_D = 1 + {1 \over 2} \sum_{k=1}^\infty {1 \over 2 k f} \left[ \ln \left( {2k-1 \over 2k+1} \right) + \ln \left( {4 f k + 2 k + 1 \over 4 f k + 2 k - 1} \right) \right].
\label{eq:bias-uniform}
\eq
The solid curve shows the bias in the neutral fraction measurement for such a model.  It is typically about half that of the model with fixed path lengths, although the two converge at small $\bxhi$.  The bias is smaller in this case because the shorter neutral path lengths decrease the apparent absorption by a larger factor than longer segments increase it, and the absorption is weighted more heavily to nearby gas.  

Next, we take a model where the ionized patches have a fixed size $R_b$ but draw the length of each neutral patch from an exponential distribution, with expectation value $f R_b$.  In this case, the probability distribution of $u=R_{ek}/R_b$ is
\bqa
p_{ek}(u) & = & {f^{-k} \over (k-1)!} \frac{\exp \left\{ - f^{-1} [u - (k-1/2)] \right\}}{u - (k-1/2)]^{-1}}, \nonumber \\ 
& & u > (k-1/2),
\label{eq:prnk}
\eqa
and zero elsewhere.  The minimum of $u$ is set by the $(k-1)$ ionized bubbles that precede this edge, plus the $R_b/2$ region immediately surrounding the source.  Because the ionized patches have fixed size, $p_{b(k+1)}(u) = p_{ek}(u-1)$.  

The long-dashed line in Figure~\ref{fig:bias-toy} shows the resulting bias.  Again, it is much reduced from the case with fixed neutral patch sizes and is similar in magnitude to the uniformly distributed case.  However, the bias in this model tends to be larger at small $\bxhi$ and smaller at large $\bxhi$.  This is because the long tail of thick absorbers allowed in the exponential model is quite significant when most absorbers are narrow.  However, when most absorbers are thick (at high $\bxhi$), the larger probability of narrow absorbers tends to decrease the bias.

Our final toy model allows both the sizes of the neutral patches and the ionized patches to be exponentially distributed, with expectation values $f R_b$ and $R_b$, respectively.  In this case, the distributions $p_{ek}(u)$ can also be written analytically, but there is no simple pattern with $k$ as in equation~(\ref{eq:prnk}).  We therefore simply present the resulting bias as the dot-dashed curve in Figure~\ref{fig:bias-toy}.  It is nearly the same as the model with fixed ionized patch size at $\bxhi \ga 0.6$:  in this regime, the neutral patches are much thicker and so their scatter dominates.  On the other hand, at small neutral fractions, the bias is much smaller in this model, because the ionized bubbles become on average larger than their neutral neighbors.\footnote{Note that the bias in this model appears to become negative at $\bxhi \la 0.04$; however, this is a numerical artifact of our approximations.}

The different biases between our toy models illustrated in Figure~\ref{fig:bias-toy} show that the bias does carry some information on the distribution of neutral and ionized gas.  In principle, an independent measurement of $\bxhi$ would then allow additional constraints on the reionization morphology.  However, this is likely to be a difficult game, because the differences are modest in the realistic models.

%%%%%%%%%%%%%%%%% FIGURE 2
\begin{figure}
\begin{center}
\resizebox{8cm}{!}{\includegraphics{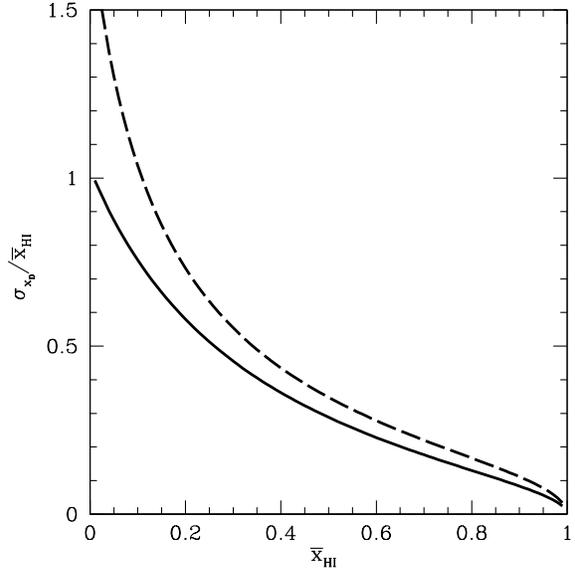}}\\%
\end{center}
%\plotone{scatplot.eps}
\caption{Fractional scatter in neutral fraction measurements using the damping wing, including only the contribution from the \emph{first} neutral region.  The solid and long-dashed curves assume that the width of the region is uniformly and exponentially distributed, respectively (as in Fig.~\ref{fig:bias-toy}).}
\label{fig:scatter-toy}
\end{figure}

\subsection{Scatter in the Measurements} \label{scatter}

Our last three models draw path lengths from different distributions; in addition to the bias, they will also therefore have scatter intrinsic to the measurement of $\bar{x}_D$.  This is more difficult to estimate analytically, because there is significant covariance between the locations of the different neutral patches (even without covariance in their individual lengths, the $i$th neutral patch must occur before the $[i+1]$th).  For a simple estimate, however, we note that most of the absorption (on average $\sim 80\%$ in our toy models) comes from the first region, so we take the variance of the first term in the sum in equation~(\ref{eq:xhd-approx}).  Again, more sophisticated treatments are possible but not necessary given our access to simulations that include many more effects than we can hope to add to analytic models.

Figure~\ref{fig:scatter-toy} shows the standard deviation in these measurements, normalized to the true ionized fraction, for two of our models from \S \ref{bias}.  The solid line assumes that the sizes of the ionized patches are fixed but that the sizes of the uniform patches are uniformly distributed; the short-dashed line assumes that the latter are exponentially distributed.  Here we see that the fractional scatter increases as $\bxhi \rightarrow 0$ (although, just as with the bias, the actual value of $\sigma_{x_{\rm HI}}$ peaks at $\bxhi \sim 0.5$).  

Interestingly, the scatter is larger for the exponentially distributed model -- unlike the bias.  This is because of the long tail allowed by the exponential distribution which becomes especially important at small $\bxhi$:  the variance of our uniform distribution is $f^2 R_b^2/3$, while the variance of the exponential distribution is $f^2 R_b^2$.  Obviously, interpreting any observations in detail will require careful modeling of the underlying distribution.

As with the bias, the scatter evolves throughout reionization.  However, it can be measured even without an independent estimate of $\bxhi$ and so can itself be used for further constraints:  a large dispersion in the measured $\bar{x}_D$ is indicative of the final stages of reionization, for example.

%%%%%%%%%%%%%%%%% FIGURE 3
\begin{figure}
\begin{center}
\resizebox{8cm}{!}{\includegraphics{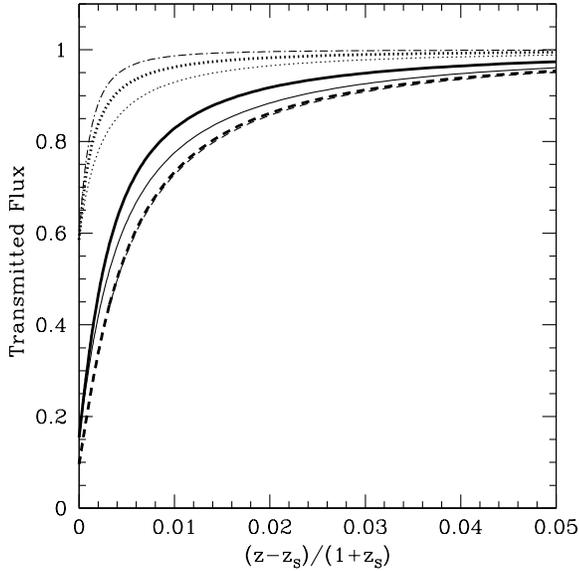}}\\%
\end{center}
%\plotone{profiles.eps}
\caption{Damping wing absorption profiles, as a function of fractional wavelength offset from the source (at redshift $z_S$).  The thick curves show the ``true" absorption profiles for $\bxhi=0.9,\,0.5$, and $0.1$ (for the dashed, solid, and dotted curves, respectively).  Note that the two dashed curves overlap and are practically indistinguishable.  The corresponding thin curves show the absorption profiles for uniformly ionized IGM normalized to the same transmission at $z_S$.  The dot-dashed curve shows the profile of a DLA, normalized to the same transmission as the $\bxhi=0.1$ curves at $z_S$.}
\label{fig:profiles}
\end{figure}

\subsection{The Absorption Profile} \label{profile}

The final question we can address with our toy model is how the ``picket fence" absorption characteristic of inhomogeneous reionization affects the damping wing absorption profile as a function of wavelength.  Of course, in our more realistic models that allow a range of absorber sizes there will be a corresponding range of profiles, and with the large number of absorbers that are allowed it is not obvious how one would fit the results or even parameterize the possibilities.  We therefore focus on the simplest model, in which the ionized and neutral regions have fixed sizes $R_b$ and $f R_b$, respectively.

Figure~\ref{fig:profiles} shows several example profiles as a function of the fractional wavelength offset from the source \lya line center (at a redshift $z_S$).  The thick curves are taken from our toy model; the dashed, solid, and dotted curves take $\bxhi=0.9,\,0.5,$ and $0.1$, respectively.  The thin curves show the corresponding profiles for a uniformly ionized IGM, beginning the same distance from the galaxy, and with an assumed neutral fraction $\bar{x}_D$.  Thus, they are normalized to have the same transmission as the ``true" curves at the redshift of the galaxy.

The profiles are nearly identical when $\bxhi$ is large, but in the middle and end stages of reionization the toy model has steeper absorption than a uniformly ionized IGM, allowing slightly more transmission redward of the source wavelength.  This is not surprising:  as described above, the gentle decline of the damping wing occurs because longer columns contribute to the absorption of photons far to the red.  In the picket-fence picture, photons far from $z_S$ are sensitive to such a large column that they average over many ionized patches.  This also explains why the effect becomes more and more important at smaller $\bxhi$, as more and more of the relevant absorbing gas is absent when the neutral patches are narrow and widely separated.

%%%%%%%%%%%%%%%%% FIGURE 4
\begin{figure}
\begin{center}
\resizebox{8cm}{!}{\includegraphics{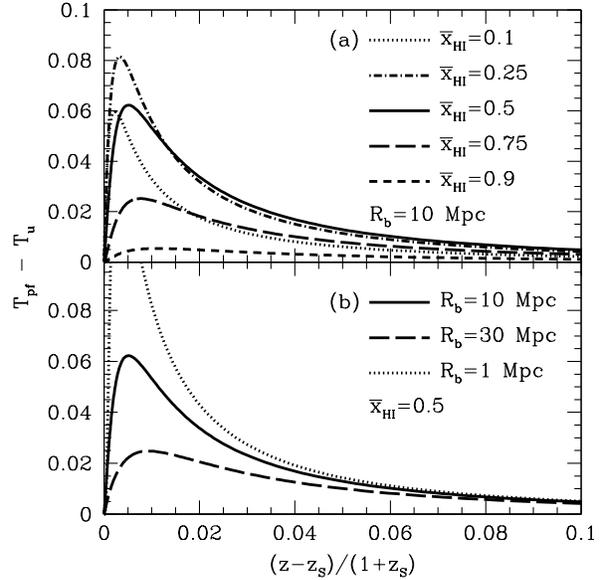}}\\%
\end{center}
%\plotone{profile_residuals.eps}
\caption{Residuals between ``true" absorption profiles and those from a uniformly ionized medium with $\bar{x}_D$, as a function of wavelength offset from the source.  In \emph{(a)}, we set $R_b=10 \Mpc$ and vary the IGM ionized fractions.   In \emph{(b)}, we fix $\bxhi=0.5$ and vary the bubble sizes. }
\label{fig:residuals}
\end{figure}

One possible worry is confusion of the damping-wing absorption with DLAs in the host galaxies of GRBs; separating the two sources of absorption requires that they have significantly different profiles (e.g., \citealt{totani06}).  The dot-dashed curve in Figure~\ref{fig:profiles} shows a DLA profile with $\tau(z_s)$ normalized to the optical depth in our picket-fence model with $\bxhi=0.1$, where the true profile is steepest.  In this case, the picket-fence model is about midway between the DLA and IGM expectations, so we would expect that clearly separating IGM and DLA absorption will become significantly more difficult toward the end of reionization.  However, when $\bxhi \ga 0.25$, the picket-fence absorption is \emph{much} closer to the naively expected IGM behavior than to the DLA behavior.
(Of course, if the DLA is centered at $z_S$, it will obscure much more of the line profile -- but this shows that they can be distinguished, at least in principle.)

Figure~\ref{fig:residuals} shows the differences in more detail.  We plot $T_{\rm pf}-T_{u}$, where $T_{\rm pf}$ is the transmission in the picket-fence model and $T_u$ is the transmission for a uniformly ionized IGM, normalized to the same optical depth at $z_S$.  Panel \emph{(a)} shows the residuals for several different ionized fractions with $R_b$ held constant (as in Fig.~\ref{fig:profiles}), while panel \emph{(b)} varies $R_b$ while holding $\bxhi$ constant.  

The deviations are typically at most a few percent, with the biggest differences when $(z-z_S)/(1+z_S) \la 0.02$.  Of course, this is also the region  most likely to be contaminated by an absorber in the host galaxy, so it is not clear how well this part of the deviation can be detected.  At redder wavelengths, the differences are $\la 2\%$, so they will require extremely high signal-to-noise spectra to detect them cleanly.

The differences at small offsets from $z_S$ are much larger for smaller bubbles, even though the bias in the estimated neutral fraction is nearly independent of $R_b$ (see Fig.~\ref{fig:bias-toy}).  This is because, when the first neutral patch is large, most of these photons are efficiently absorbed inside of it.  When the patch is small, the effective column decreases relatively rapidly. 
Note that the strongest differences in the profiles, especially at small wavelength offsets, are due to $R_b$.  Thus most of the variations in the spectral shape will help to pin down the bubble size (which we have assumed can be measured independently).  This suggests that it will be difficult to use variations in the shape to measure $\bar{x}_D$ more accurately, at least not in any straightforward fashion.

It is obvious from this section that the damping wing spectra contain more information than just the location of the nearest neutral gas, $R_{b1}$, and $\bar{x}_D$ (which we have assumed to be the two observables).  However, it is not clear whether higher-order differences can be measured in practice, because of the finite signal-to-noise to be expected from these sources and because of intervening absorption in the host galaxy itself.  This is especially true because the number of extra parameters required is large:  for example, we have found that including only the first neutral region leads to residuals of similar magnitude to those in Figure~\ref{fig:residuals} (though with the opposite sign, because such a procedure always underestimates the total amount of absorption).  Thus an accurate fit would require adding the contributions from many neutral patches, each with unknown location and width.

Moreover, there will of course be scatter in the profiles at a given $(\bar{x}_D,\,R_{b1})$ because of scatter in the sizes of ionized and neutral patches.  In the following we will therefore take the simple-minded viewpoint that only these two quantities can be measured, although we will use our simulations to measure the dispersion in the profiles.  We defer a more detailed investigation of parameter estimation to future work.

\section{Semi-Numerical Simulations}
\label{sim}

In the remainder of the paper, we will use more reliable and physically-motivated calculations of inhomogeneous reionization that incorporate the full geometry of the IGM to examine the same issues of bias and scatter in damping wing measurements.  We use an excursion-set approach combined with first-order Lagrangian perturbation theory to efficiently generate density, velocity, halo, and ionization fields at $z=9$.  This ``semi-numerical'' simulation is presented in \citet{mesinger07}, to which we refer the reader for details.  A similar halo-finding scheme has also been presented by \citet{bond96-algo} and a similar scheme to generate ionization fields has been presented by \citet{zahn07-comp}.

Our simulation box is 250 Mpc on a side, with the final density, velocity and ionization fields having grid cell sizes of 0.5 Mpc.  Halos with a total mass $M \ge 2.2 \times 10^8$ $\Msun$ are filtered out of the linear density field using excursion-set theory, with mass scales spaced as $\Delta M / M = 1.2$.
Note that we are able to resolve halos with masses less than a factor of two from the cooling mass likely to be pertinent mid-reionization, corresponding to gas with a temperature of $T\sim10^4$ K (e.g. \citealt{efstathiou92, thoul96, gnedin00-fb, shapiro94}).  Halo locations are then adjusted using first-order Lagrangian perturbation theory.  The resulting halo field matches both the mass function and statistical clustering properties of halos in N-body simulations \citep{mesinger07}.

In constructing the ionization field, the IGM is modeled as a two-phase medium, comprised of fully ionized and fully neutral regions (this is a fairly accurate assumption in the context of damping wing absorption before the end of reionization, unless the X-ray background is rather strong).  Using the same halo field at $z=9$, we generate ionization fields corresponding to different values of $\bxhi$ by varying a single efficiency parameter, $\zeta$, again using the excursion-set approach (c.f. \citealt{mesinger07, furl04-bub}).  

This semi-numeric method is thus ideally suited to the damping wing problem, because we are able to ``resolve" relatively small halos and simultaneously sample a large, representative volume of ionized bubbles.  

Our procedure for computing $\tau_D$ is fully described in \citet{mesinger07-lya}, with a couple of minor differences noted below; note that similar results were also obtained by \citet{mcquinn07-lya} using full radiative transfer simulations.  We use equation~(\ref{eq:jordi_tau}) to calculate the \lya optical depth for each neutral hydrogen patch, summing the contributions of patches along the LOS for 200 Mpc or until the first neutral patch is encountered, whichever comes last,\footnote{This number was chosen experimentally in order to ensure convergence of the $\taudamp$ distributions at the mass scales and neutral fractions studied in this work.} wrapping around the simulation box if needed.  We construct distributions of $\taudamp$ for halo mass scales in the range $2.5\times10^{9}$ -- $2.4\times10^{10}$ $\Msun$ and for various ionization topologies (i.e. $\bxhi$).  We make sure to process LOSs from every halo of a particular mass scale, cycling through the halo list until each mass scale undergoes a minimum of $3\times10^4$ such Monte Carlo realizations.  Unlike \citet{mesinger07-lya}, we do not include the peculiar velocities of halos in estimating $\taudamp$.\footnote{Ignoring velocities simplifies the analysis, since we are guaranteed not to have halos whose Ly$\alpha$ line centers have Doppler shifted into resonance in the neutral IGM.  For such objects, the damping wing could still be recovered by looking further redward in their spectra.  The same fundamental quantities (especially an analog of $\bar{x}_D$) could still be measured from such sources, but they would need to be re-parameterized.  For the purposes of our statistical analysis it is useful to compare absorption at a single redshift offset across all objects, which we take to be the line center, $z_S$.  We have verified that including halo peculiar velocities mainly serves to increase the scatter at high neutral fractions, when HII bubbles are small, as expected from the preceding argument.  The inclusion of velocities is uncertain in any case because we ignore the possibility of galactic winds and correlations of the velocity field on large scales.}

\section{Results}
\label{results}

\begin{figure}
\begin{center}
\resizebox{8cm}{!}{\includegraphics{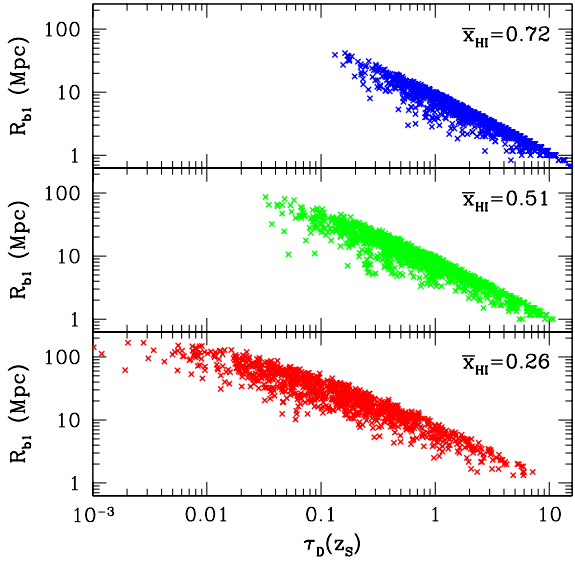}}\\%
\end{center}
%\plotone{log_tau_vs_R.eps}
\caption{Scatter plot of the distance to the nearest neutral clump, $R_{b1}$, as a function of the damping wing optical depth, for several different phases of reionization: $\bxhi =$ 0.72, 0.51, 0.26 (top to bottom).  Each panel was created using 1000 randomly chosen LOSs. The large scatter in $R_{b1}$ at a fixed $\tau_D(z_S)$ [or in $\tau_D(z_S)$ at a fixed $R_{b1}$] is one source of the difficulties with damping wing measurements.  }
\label{fig:tau_vs_R}
\end{figure}

In this section, we use the semi-numeric simulations to revisit the issues of bias and scatter raised in \S \ref{toy}.  We begin by illustrating the difficulty involved in accurately estimating $\bxhi$ in an inhomogeneously ionized IGM.  We have already seen that there is a {\it deterministic} and accurate mapping $(R_{b1}, \tau_D) \leftrightarrow \bar{x}_D = \bxhi$ (c.f.~eq.~\ref{eq:xhd}) in a uniformly ionized IGM.  However, as discussed previously, under the more realistic picket-fence IGM absorption scenario, this mapping becomes {\it stochastic}.  To illustrate this, in Figure \ref{fig:tau_vs_R}, we show a scatter plot of the distance to the nearest neutral clump, $R_{b1}$, as a function of the damping wing optical depth.  Each panel was created using 1000 randomly chosen LOSs in several different phases of reionization:  $\bxhi =$ 0.72, 0.51, 0.26 (top to bottom).  The large scatter in $R_{b1}$ at some fixed $\tau_D(z_S)$ is one source of difficulties with damping wing measurements:  it implies that LOSs with identical apparent optical depths (at a specified wavelength) pass through very different patterns of HI.  As the analytic models predicted in the previous section, this scatter increases with decreasing $\bxhi$.

%%%%%%%%%%%%%%%%% FIGURE 5

%%%%%%%%%%%%%%%%% FIGURE 6
\begin{figure}
\begin{center}
\resizebox{8cm}{!}{\includegraphics{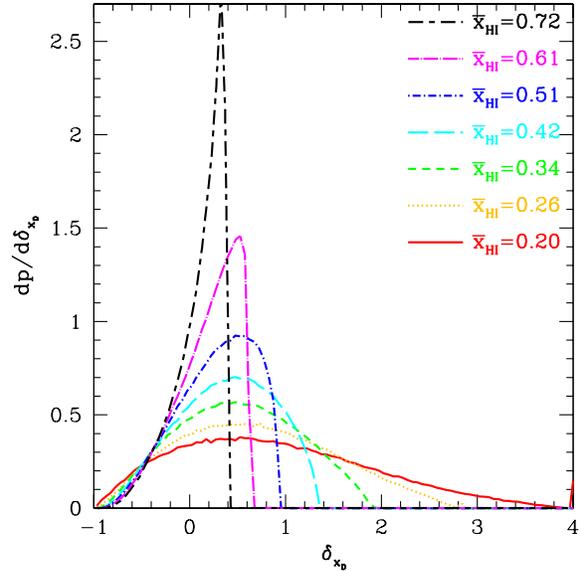}}\\%
\end{center}
%\plotone{hist.eps}
\caption{
Probability distribution of $\delta_{x_D} \equiv (\bar{x}_D/\bxhi-1)$ for several different phases of reionization: $\bxhi =$ 0.20, 0.26, 0.34, 0.42, 0.51, 0.61, 0.72 (right to left at large $\delta_{x_D}$).  Note that the mean of $\delta_{x_D}$ is non-zero, and that the distribution becomes wider and more biased as reionization progresses.
}
\label{fig:hist}
\end{figure}

\subsection{The Distribution of the Inferred Neutral Fraction} \label{bias-sim}

In Figure \ref{fig:hist}, we plot the probability distribution of $\delta_{x_D} \equiv (\bar{x}_D/\bxhi-1)$ for several different phases of reionization: $\bxhi =$ 0.20, 0.26, 0.34, 0.42, 0.51, 0.61, 0.72 (right to left at large $\delta_{x_D}$).  The mean of $\delta_{x_D}$ is greater than zero for all phases, as predicted by the toy models in the previous section: $\bar{x}_D > \bxhi$.  In addition, the bias and scatter increase as reionization progresses.  Note also that the probability distributions of $\delta_{x_D}$ are highly non-gaussian due to the restricted range  $0 \leq \bar{x}_D \leq 1$.  

The small spike at $\delta_{x_D}=4$ in the $\bxhi=0.2$ distribution results from LOSs containing a single neutral patch at the end of their path length, to which our prescription assigns $\bar{x}_D=1$.  In reality, the optical depth along such LOSs is so small that $\tau_D$ (and especially $R_{b1}$) would probably not be measurable in the first place, and in any case our assumptions of fully ionized bubbles and a constant $\bxhi$ along each geodesic break down in this regime.  However such LOSs comprise less than 0.7\% of the total at $\bxhi = 0.2$, the smallest neutral fraction we study, and so do not significantly affect the statistical estimates below.  We do note that the abundance of these LOSs does depend on our semi-numeric algorithm.  In this example, the algorithm of \citet{zahn07-comp} has 14\% of LOSs in this regime at $\bxhi=0.2$.  However, if these unusual LOSs (which are, as we have argued, probably useless for this measurement) are removed from the sample, both algorithms agree on the net bias.

\begin{figure}
\begin{center}
\resizebox{8cm}{!}{\includegraphics{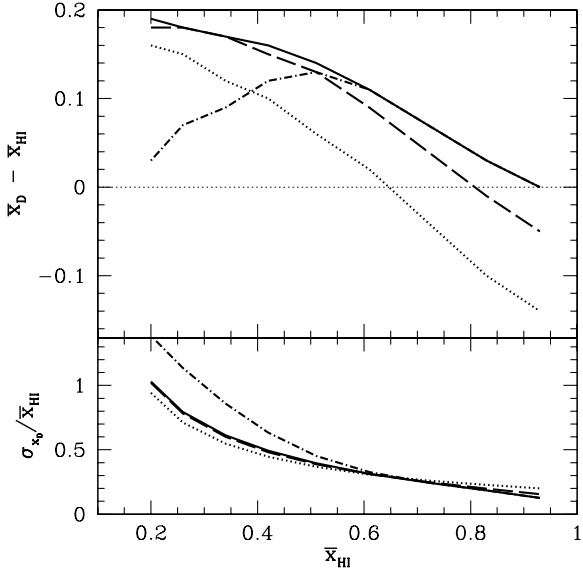}}\\%
\end{center}
%\plotone{mean_sig.eps}
\caption{\emph{Top:}
The damping wing bias from assuming a constant neutral fraction throughout the IGM, expressed as $(\bar{x}_D - \bxhi)$, as in Fig. \ref{fig:bias-toy}.
\emph{Bottom:}  Fractional scatter in $\bar{x}_D$.
In both panels, the solid curve is generated from all LOSs, the dashed (dotted) curve is generated from LOSs with $\tau_D(z_S) <$ 5 (1), and the dot-dashed curve is generated assuming $\bar{x}_D = 0$ for LOSs with $\tau_D(z_S) < 0.1$.  Note that the total mean bias is always positive, but removing parts of the distribution to mimic real observational data sets can drive the apparent mean bias to negative values at large $\bxhi$.
}
\label{fig:mean_sig}
\end{figure}

In Figure \ref{fig:mean_sig}, we plot the bias expressed as $(\bar{x}_D - \bxhi)$ (top panel), and the fractional scatter in $\bar{x}_D$ (bottom panel).  The solid curve is generated from all of the LOSs.  This net bias is always positive and matches our toy model with exponentially distributed neutral patches fairly well.  However, we do not see evidence for a turnover at small $\bxhi$, and the simulation curve also increases somewhat more slowly then the toy model.

The first of these differences has a simple explanation.  Weak damping wing absorption might not be detectable with finite signal-to-noise observations (or it may not be separable from an uncertain source continuum), and the corresponding sources would likely be labeled as post-reionization objects.  Thus, the dot-dashed curve is generated by setting $\bar{x}_D = 0$ for LOSs with $\tau_D(z_S) < 0.1$ (see \citealt{mesinger07-lya} for the total optical depth distributions).  Imposing a minimum value of $\tau_D$ imposes a minimum on $\bar{x}_D$, so the bias starts decreasing at low $\bxhi$.  As the neutral fraction decreases, the number of these LOSs increases rapidly; eventually, the apparent distribution will divide into a large set of apparently absorption-free spectra and a few spectra where the inferred neutral fraction is large.  Of course, both sets must be taken into account to yield the strongest constraints. 

We also note that the absolute value of the bias is not particularly important, so long as it can be calibrated through simulations like this one.  The crucial point will be to understand the sample and the model well enough to perform this calibration; otherwise systematic uncertainties will remain in the interpretation of the observations.  As noted above, the assumptions inherent in the particular radiative transfer or semi-numerical algorithm used to generate the ionization field become increasingly important at $\bxhi\lsim0.2$; thus we predict that damping wing measurements in a $\bxhi\lsim0.1$ universe will be very difficult to interpret.  We have not extended our models to this regime because our semi-numerical algorithm is no longer very robust (as the comparison to the \citealt{zahn07-comp} algorithm above shows) and because evolution across the line of sight will play an increasingly important role; a ``light-cone" analysis will probably be required for such a regime.

On the other hand, real-world instruments are not infinitely sensitive, so samples generated by Ly$\alpha$ emission line searches (e.g., narrow band surveys) might only contain objects whose optical depth is less than some maximum value following from the instrumental sensitivity.  To model this possibility, the dashed  and dotted curves in Figure~\ref{fig:mean_sig} are generated only from LOSs with $\tau_D(z_S) <$ 5 and $1$, respectively.  Note that the apparent bias calculated from such truncated distributions becomes {\it negative} at large $\bxhi$, because imposing a maximum $\tau_D$ also imposes a maximum on $\bar{x}_D$ (for a given $R_{b1}$).  

The fractional scatter is comparable for all the curves.  It increases most steeply at low values of $\bxhi$ for the dot-dashed curve, because setting $\bar{x}_D = 0$ for LOSs with weak absorption induces a bi-modal distribution of $\bar{x}_D$, with small values of $\bar{x}_D$ shifting to a spike at $\bar{x}_D=0$ (c.f. Fig. \ref{fig:hist}).  Again, the scatter can be calibrated with these types of models, but it requires fairly large samples to interpret the red damping wing reliably.

%%%%%%%%%%%%%%%%% FIGURE 7

\subsection{The Damping Wing Profile} \label{prof-sim}

In \S \ref{profile} we noted that our toy models of picket-fence absorption produce a steeper absorption profile than expected in a homogeneously ionized IGM.  Here we confirm and further quantify this result using our semi-numerical simulations.  Specifically, we parametrize the absorption profile as (c.f. eq. \ref{eq:tau-approx}):
\bq
\label{eq:profile_fit}
\tau_D(z) \propto R_{b1}^{-\alpha} ~ .
\eq
Note from eq. (\ref{eq:tau-approx}) that in the homogenously ionized IGM, $\alpha = 1$; however, from Fig. \ref{fig:profiles}, we expect that during patchy reionization the mean power law index $\bar{\alpha}$ is greater than unity.

\begin{figure}
\begin{center}
\resizebox{8cm}{!}{\includegraphics{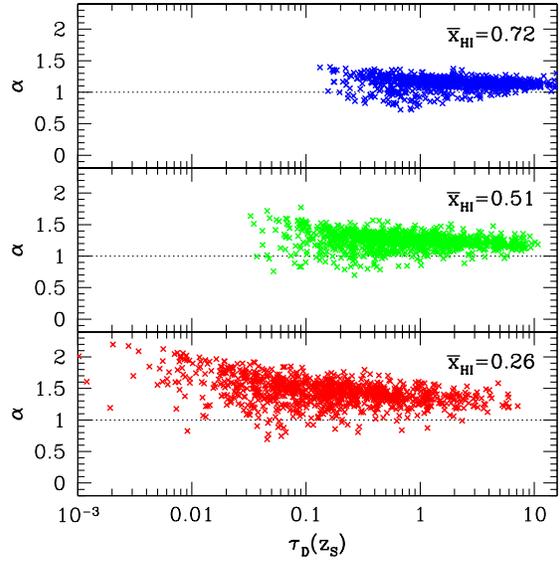}}\\%
\end{center}
%\plotone{profile.eps}
\caption{
Scatter plot of the power law index $\alpha$ from the damping wing profile parameterization in eq. (\ref{eq:profile_fit}), fit using two points at $z=z_S$ and $z=z_S+0.1$.  Panels assume $\bxhi =$ 0.72, 0.51, 0.26 (top to bottom).  The horizontal lines denote $\alpha = 1$, as would be expected from a uniformly ionized IGM.
}
\label{fig:profile}
\end{figure}

To test this with our simulations, we perform a simple two point power law fit to the profile shape, calculating $\taudamp(z_S)$ and $\taudamp(z_S+0.1)$.  A scatter plot of the resulting power law index, $\alpha$, from 1000 randomly chosen LOSs is shown in Figure \ref{fig:profile}. Panels assume $\bxhi =$ 0.72, 0.51, 0.26 (top to bottom).  It is obvious from the figure that indeed $\bar{\alpha} > 1$, with the profile steepening and the scatter increasing as reionization progresses.  Note also that the mean and scatter of $\alpha$ change with $\bxhi$, {\it even at fixed $\taudamp(z_S)$}.  Although LOSs can have similar integrated HI columns lengths at different epochs of reionization, the distribution of neutral hydrogen along this subset of LOSs  does evolve.  In general, LOSs intersect {\it a fewer number of longer} neutral patches at high $\bxhi$ than LOSs with the same $\taudamp$ at low $\bxhi$.  Shorter neutral patches, especially those close to $R_{b1}$, translate in turn into a steeper absorption profile (see the discussion in \S \ref{profile}).  

%%%%%%%%%%%%%%%%% FIGURE 8

Of course, the systematic variation of $\alpha$ with $\bxhi$ implies that the spectral shape can also be used to constrain the latter.  However, because the dispersion in profiles is always at least as large as the differences in the means (even at fixed $\tau_D$), it will require large samples to take advantage of this information.

%%%%%%%%%%%%%%%%% FIGURE 9
\begin{figure}
\begin{center}
\resizebox{8cm}{!}{\includegraphics{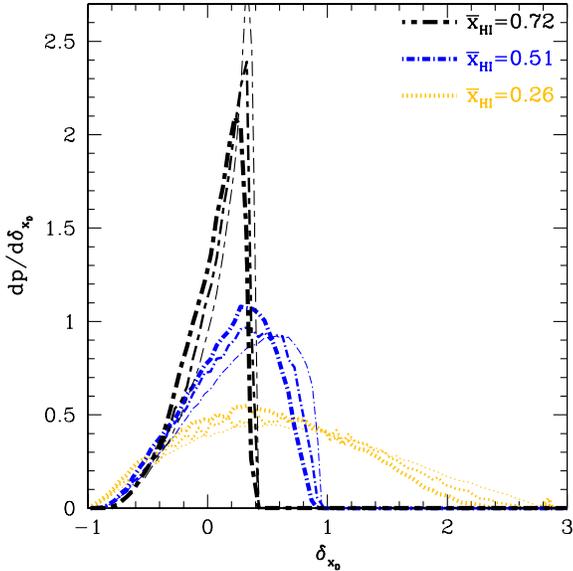}}\\%
\end{center}
%\plotone{mass_hist.eps}
\caption{
Probability distributions of $\delta_{x_D} \equiv (\bar{x}_D/\bxhi-1)$ generated from LOSs originating from halos with masses $M =$ 2.6$\times10^{11}$,  2.5$\times10^{10}$, and 2.3$\times10^9$ $\Msun$ (thick to thin curves), at several different phases of reionization: $\bxhi =$ 0.26, 0.51, 0.72 (right to left at large $\delta_{x_D}$).
}
\label{fig:mass_hist}
\end{figure}

\subsection{Variation with Halo Mass} \label{halomass}

\citet{mesinger07-lya} showed that the mean and dispersion of $\taudamp(z_S)$ were functions of halo mass at a fixed neutral fraction.  Most of this variation is due to differences in the HII bubbles that these halos reside in (because larger halos are more clustered and so tend to sit in larger bubbles; \citealt{furl04-lya}).  But any excess difference would lead to another bias in the interpretation of damping wings.

In Figure \ref{fig:mass_hist}, we plot the probability distributions of $\delta_{x_D} \equiv (\bar{x}_D/\bxhi-1)$ generated from LOSs originating in halos with masses $M =$ 2.6$\times10^{11}$,  2.5$\times10^{10}$, and 2.3$\times10^9$ $\Msun$ (thick to thin curves), at several different phases of reionization: $\bxhi =$ 0.26, 0.51, 0.72 (right to left at large $\delta_{x_D}$).  The $M =$ 2.6$\times10^{11}$ $\Msun$ halos are the largest halos in our simulation at $z=9$, with our 250 Mpc box containing eight of them.

There are clearly some differences in the inferred values of $\bar{x}_D$ from the different types of halos:  LOSs originating from more massive halos have somewhat narrower distributions, with smaller means, at fixed $\bxhi$.  This is because more massive halos generally sit inside larger bubbles (with larger $R_{b1}$) and so the Ly$\alpha$ absorption cross-section is flatter when photons enter their first neutral patch.  As discussed previously, it is the varying cross-section that causes the bias and scatter in measurements; if the Ly$\alpha$ cross-section were completely flat, $\bar{x}_D$ would always equal $\bxhi$.

\begin{figure}
\begin{center}
\resizebox{8cm}{!}{\includegraphics{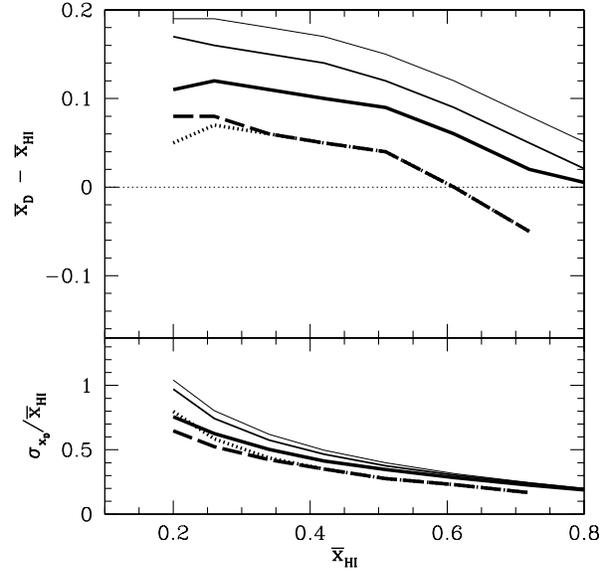}}\\%
\end{center}
%\plotone{high_mass_mean_sig.eps}
\caption{
Damping wing bias statistics.  Solid lines were created using all LOSs originating from halos with masses $M =$ 2.6$\times10^{11}$,  2.5$\times10^{10}$, and 2.3$\times10^9$ $\Msun$ (thick to thin). Dashed and dotted lines were created using only LOSs originating from halos with masses $M =$ 2.6$\times10^{11} \Msun$ {\it and} with $R_{b1} > 40$ Mpc; the dotted lines additionally assume that $\bar{x}_D = 0$ for LOSs with $\tau_D(z_S) < 0.01$.
\emph{Top:} The damping wing bias from assuming a constant neutral fraction throughout the IGM, expressed in $(\bar{x}_D - \bxhi)$.
\emph{Bottom:}  Fractional scatter in $\bar{x}_D$.
}
\label{fig:high_mass_mean_sig}
\end{figure}

However, Figure~\ref{fig:mass_hist} also shows that the distributions are much more sensitive to $\bxhi$ than to the halo mass and hence can be robustly used to estimate $\bxhi$ from observational data sets even with little or no knowledge about the underlying halo.  But the overall bias does have a non-negligible dependence on mass, so such information will be useful.  We quantify this in Figure~\ref{fig:high_mass_mean_sig}, where the solid curves correspond to the same three mass scales as in Figure~\ref{fig:mass_hist}: $M =$ 2.6$\times10^{11}$,  2.5$\times10^{10}$, and 2.3$\times10^9$ $\Msun$ (thick to thin, or bottom to top).  The mean bias, $\bar{x}_D - \bxhi$, increases by a factor of $\sim$2 as the host mass scale is decreased from 2.6$\times10^{11}$ to 2.3$\times10^9$ $\Msun$, and the scatter at small $\bxhi$ also decreases somewhat.  Thus, if the properties of the host halo can be measured, it will certainly help to extract stronger constraints.  In the following subsection, we examine such a special case.

%%%%%%%%%%%%%%%%% FIGURE 10

\subsection{Quasars}
\label{sec:qsos}

As mentioned in the introduction, \citet{mesinger04, mesinger07-prox} already claim to have detected 
damping wings in two high-redshift quasars.  Their model assumed a uniform UV background flux for the purposes of calculating the damping wing, so their final constraint is comparable in spirit to our $\bar{x}_D$ parameter, 
though it is not clear if their results are dominated by the damping wing profile shape or the inferred $\tau_D$ (and likewise $x_D$), the later being partially degenerate with other free parameters in the analysis.

Obviously, it would be interesting to study the effectiveness of such damping wing constraints when an inhomogeneously ionized IGM is included, even more so considering that they can be applied to future high-redshift data sets.  Note that, although our semi-numerical simulation boxes are at $z_S=9$, well beyond the SDSS (and possibly future) quasars, the ionization topology and optical depth statistics are weak functions of redshift in this range \citep{mcquinn07, mesinger07-lya}.

Unlike normal galaxies, which are the focus of most of this work, bright quasars lie in highly biased regions with correspondingly large $R_{b1}$.  We have already seen in Figures~\ref{fig:mass_hist} and \ref{fig:high_mass_mean_sig} that this decreases the scatter and bias. In Figure \ref{fig:high_mass_mean_sig}, the thick curves show the bias and scatter computed from LOSs originating from the most massive halos in our simulation box, $M = 2.6 \times 10^{11} \Msun$.  For the present analysis, where the bright quasar necessarily produces a large HII region, we are interested only in rare LOSs with large $R_{b1}$.  To guarantee convergence of our measurements in these unusual cases, we extend our path length of integration to 400 Mpc, although we find that this only has a noticeable effect for the $\bxhi=0.2$ data point.   The thick solid line was created using all LOSs.  The dashed lines were created using the subset of LOSs with $R_{b1} > 40$ Mpc, typical of the high-$z$ SDSS quasars.\footnote{LOSs with such large $R_{b1}$ are very rare at high $\bxhi$ and our box only contains them when $\bxhi \lsim 0.75$.}  

Both the bias and the scatter decrease compared with the solid curve.  Requiring that $R_{b1}$ be large means that the Ly$\alpha$ absorption cross-section at $R_{b1}$ is flatter than usual, with $(\bar{x}_D - \bxhi)$ smaller by $\sim 0.05$ throughout.  
We caution however that the curves in Figure \ref{fig:high_mass_mean_sig} are calculated at $z_S$; thus if one is estimating $x_D$ blueward of the line center (i.e. using $\tau_D(z < z_S)$), the bias is likely to lie somewhere between the solid and dashed curves.
Note also that the bias shown with the dashed curve becomes {\it negative} at $\bxhi \gsim 0.6$ (see Fig.~\ref{fig:mean_sig} and discussion thereof).  The dotted lines in Fig. \ref{fig:high_mass_mean_sig} were also created using LOSs with $R_{b1} > 40$ Mpc, but with the additional assumption that $\bar{x}_D = 0$ for LOSs with $\tau_D(z_S) < 0.01$.\footnote{Note that we assume the damping wing is studied blueward of the Ly$\alpha$ line center for quasars, so its footprint can be non-negligible even with a small line-center optical depth $\taudamp(z_S)$.  $\tau_D(z_S) \sim 0.01$ would roughly be expected if $\bxhi\sim0.1$ and $R_{b1} \sim 40$ Mpc, as in the observed systems.}  Having an effective minimum $\taudamp$ results in the same decrease of bias and increase in scatter as was seen in Figure~\ref{fig:mean_sig}.

\section{Discussion}
\label{disc}

In this paper, we have examined how the shape of the \lya red damping wing can be used to constrain the IGM before reionization is complete.  In the past, it has usually been assumed that the absorbing gas can be well-approximated by a uniform density medium with constant ionized fraction.  However, recent reionization models have shown that ionized bubbles can be quite large, so the latter is not a good approximation.  We have therefore critically examined how well the damping wing constrains the neutral fraction during inhomogeneous reionization.

We have identified two major issues with its interpretation.  First, there is substantial scatter in the optical depth along different lines of sight.  Most of this is due to the scatter in the distance between the source and the nearest patch of neutral gas; however, there is still non-negligible scatter even if this distance can be measured from the shape of the damping wing.  In our semi-numeric simulations, the fractional r.m.s. fluctuation in $\bxhi$ thus estimated increases from 0.1 to 1 over the range $0.9\gsim\bxhi\gsim0.2$.   Fortunately, this statistical uncertainty can be reduced simply by finding more lines of sight.

The other problem is more severe:  we have shown that the ``picket-fence" absorption from inhomogeneous reionization adds a systematic, and often large, \emph{bias} to measurements of the neutral fraction.  Although the damping wing is indeed sensitive to a large path length through the IGM, it is most sensitive to the closest gas.  As a result, simple fits to the damping wing will always \emph{overestimate} the true neutral fraction in a partially ionized universe, with an error of $\sim 30\%$ near the middle of reionization.  This bias is generic to any inhomogeneous model.  The bias is reduced and can even become negative if observations only probe a subset of the entire halo population, such as quasars with large HII regions.

Both the systematic and statistical uncertainty can be reduced by a careful fit to the damping wing spectral profile, which is typically steeper than the naively expected $(\Delta \lobs)^{-1}$ profile.  However, because the absorption typically comes from many neutral patches, a large number of parameters are required for a detailed fit, and given the relatively modest difference from the expected behavior, these will be difficult to measure, probably only possible in systems with intrinsically large optical depths.  
Moreover, the scatter in the profiles, even at fixed $\tau_D$, is sufficient that large samples will be required to put strong constraints on reionization from the spectral shape.

Of course, the bias and scatter also depend on $\bxhi$ and so can, at least in principle, be used to constrain it.
For example, large dispersion in the inferred neutral fractions could be an indicator of $\bxhi \la 0.2$.  
If an independent estimate of $\bxhi$ exists, one could reverse the direction of analysis, and use the bias and scatter to constrain the reionization model and topology.

Fortunately, for a given model of reionization, the dispersion and bias can be calibrated by theoretical models.  We therefore argue that the most efficient way to constrain reionization with the damping wing is through comparison with detailed models.  Of course, any such constraints will be model-dependent, but we believe that the morphology of reionization is now sufficiently well-understood (see, e.g., \citealt{furl05-charsize, mcquinn07}) that these uncertainties will likely not dominate the statistical uncertainties from the small number of accessible sources, at least in the relatively near future.  For example, the reionization morphology is nearly independent of redshift \citep{furl04-bub, mcquinn07}.  
Also, we have found only a modest dependence of the $x_D$ distribution on halo mass (mostly due to the variation in bubble size with mass).  However, toward the end of reionization, when the absorption is dominated by rare, narrow sheets of neutral hydrogen, the details of the radiative transfer algorithm (or an approximation to it, as in our models) and of the sample selection will be extremely important.  Nevertheless, the task is challenging, as the damping wing profile must be separated from the rapidly varying resonance absorption for quasars (as in \citealt{mesinger04, mesinger07-prox}) or from intrinsic absorbers for GRBs.  Fortunately, in the latter case $\sim 20\%$ of moderate-redshift GRBs have only modest absorbers and will still be useful for these purposes \citep{chen07}.  

So far, the damping wing analysis has been performed on three high-redshift quasars: \qnamefourtwo\ ($z_S=6.42$), \qnametwoeight\ ($z_S=6.28$), \qnametwotwo\ ($z_S=6.22$) \citep{mesinger04, mesinger07-prox}, as well as GRB 050904 ($z_S \approx 6.3$) \citep{kawai06, totani06}.  This paper highlights the need to calibrate these and future damping wing analysis with simulations of the reionization morphology.  Obviously we cannot set firm constraints without detailed simulations of the observations.  Nevertheless, the mean bias we find from our simulations seems to work in the direction of strengthening the upper limit (on $\bxhi$) from the \citet{totani06} measurements, and weakening the lower limit from the \citet{mesinger04, mesinger07-prox} constraints at $\bxhi\lsim0.6$ (although, interestingly, it would strengthen them if $\bxhi\gsim0.6$; see the sign change for the bias in Fig.~\ref{fig:high_mass_mean_sig}).  Conversely, the steeper-than-expected absorption profile seems to work in the direction of weakening the \citet{totani06} constraints (especially because it must be distinguished from strong internal absorption) while strengthening the \citet{mesinger04, mesinger07-prox} constraints.  
The absorption profile might be more relevant than the bias for these studies, as an overall bias can be partially degenerate with other free parameters in the fit:  that is, when the absorption profile \emph{can} be detected to high precision, its shape will certainly be useful in constraining $\bxhi$.  The scatter in both effects would probably somewhat erode the confidence contours for all of these studies. 
On the other hand, our model predicts large scatter between different LOSs at the end of reionization, which is consistent with the measurements at $z \sim 6.3$.
More precise limits will require simultaneous fits to the intrinsic absorption and the range of possible IGM absorber profiles, and we defer them to future work.

Another intriguing possibility is to try to measure damping wing characteristics from stacked spectra of many \lyans-emitting galaxies.  \citet{mcquinn07} have shown that the wing shape is difficult to separate from uncertainties in the line for individual objects, and the scatter we have described will also make the interpretation of individual faint emitters problematic.  But, if the characteristics of the population are relatively constant, stacking may increase the signal to noise sufficiently to allow a detection of a ``mean" damping wing at each redshift, even far redward of line center.  

%\acknowledgments

SRF thanks Crystal Martin and Josh Bloom for conversations that stimulated this work.  We thank Z. Haiman for helpful comments on this manuscript.  This research was partially supported by grant NSF-AST-0607470. 

\bibliographystyle{mn2e}
\bibliography{Ref_21cm,Ref_2007}

\end{document}